\documentclass[reprint,amsmath,amssymb,aps]{revtex4-2}
\usepackage{graphicx}
\usepackage{bm}
\usepackage{array}
\usepackage{tabularx}
\usepackage[hidelinks]{hyperref}
\usepackage[english]{babel}

\begin{document}

\title{Parametric $\mathrm{CZ}$ Gates in a Tunable Tripod Architecture with Fixed-Frequency Superconducting Qubits}

\author{Chanpyo Kim}
\email{freezeticket@hanyang.ac.kr}
\affiliation{Department of Applied Physics, Hanyang University, Ansan, Republic of Korea}

\author{Hansol Kim}
\email{khshk18@hanyang.ac.kr}
\affiliation{Department of Applied Physics, Hanyang University, Ansan, Republic of Korea}

\author{Jeongsoo Kang}
\email{jskang@sdt.inc}
\affiliation{Teheran 44-gil 5, Gangnam-gu, Seoul, Republic of Korea}

\author{Younghun Kwon}
\email[Corresponding author: ]{yyhkwon@hanyang.ac.kr}
\affiliation{Department of Applied Physics, Hanyang University, Ansan, Republic of Korea}

\date{November 2, 2026}

\begin{abstract}
Efficient implementation of the operations required for quantum error correction in superconducting quantum computers requires high gate fidelity, adequate qubit connectivity, and selective control of interactions. Sharing a single coupler among multiple qubits can improve connectivity, but also requires accounting for its effects on qubits outside the target pair. In this work, we numerically investigate the feasibility of implementing parametric-resonance $\mathrm{CZ}$ gates in a tunable tripod structure consisting of three fixed-frequency transmon qubits connected to a single tunable coupler. We modulate the external flux applied to the coupler and optimize the pulse parameters for each qubit pair to construct $\mathrm{CZ}_{01}$, $\mathrm{CZ}_{02}$, and $\mathrm{CZ}_{12}$ gates. We calculate the time evolution of a Hamiltonian that includes three energy levels for each circuit element and evaluate gate fidelity and leakage after virtual Z correction. The fidelities obtained for the target-pair computational blocks, whose input and output basis states have both the coupler and spectator qubit in their ground states, are 0.9943, 0.9834, and 0.9907, respectively. We also find that population preservation at the end of the gate depends on the initial state of the spectator qubit. These results demonstrate the possibility of constructing $\mathrm{CZ}$ operations for each qubit pair through flux modulation of a shared coupler while maintaining fixed data-qubit frequencies. They also highlight the need to evaluate performance and optimize control over the full computational space, including the spectator qubit.

\end{abstract}

\maketitle

\section{Introduction}

Quantum computing is being explored as a means of addressing a variety of computational problems. Representative approaches, including the variational quantum eigensolver (VQE) and quantum machine learning, investigate applications to problems that are difficult to treat classically.\cite{montanaro_quantum_2016, biamonte_quantum_2017, choe2025efficient} These algorithms are implemented as quantum circuits comprising sequences of quantum gates, and the effects of gate errors and decoherence can accumulate as circuit depth increases.\cite{burnett2019decoherence} Limited connectivity can also substantially increase the number of $\mathrm{SWAP}$ gates required, thereby increasing circuit depth. Reliable quantum computation therefore requires not only high individual gate fidelity but also connectivity and control schemes that enable efficient implementation of the required operations on a given hardware platform.

As the scale of quantum computation increases, quantum error correction (QEC) becomes increasingly important for managing gate and measurement errors.\cite{gottesman_introduction_2009, devitt_quantum_2013} QEC encodes logical-qubit information into multiple physical qubits and detects and corrects errors through syndrome measurements, making it a key component of fault-tolerant quantum computing (FTQC).\cite{fowler2012surface, chow_universal_2012, kim2025implementation} Syndrome measurements require prescribed interactions between physical data and ancilla qubits. Consequently, the implementation cost of QEC is closely related to hardware connectivity and the methods used to implement gates.\cite{valentini2025demonstration}

Recently, various quantum low-density parity-check (qLDPC) codes have been proposed to enable efficient QEC using fewer physical qubits. Most of these codes employ stabilizers of weight six or higher, and implementing them requires connectivity of four or higher, compared with the degree-four connectivity required by conventional surface codes.\cite{forlivesi2024logical} In an architecture with a separate coupler for every required qubit-pair interaction, increasing the number of connections can increase the number of couplers and the complexity of circuit layout and interaction calibration. It is therefore useful to consider architectures in which multiple qubits share a coupler, providing the required connectivity while reducing circuit and control overhead.\cite{krinner2022realizing}

Architectures in which multiple qubits share a single coupler have been investigated from this perspective. Previous studies connected three or four fixed-frequency transmon qubits to a single fixed-frequency coupler and implemented $\mathrm{CX}$ gates through microwave-driven cross-resonance (CR) interactions.\cite{magesan_effective_2020, kang_new_2025, kim_lattice_2026} This approach implements gates using microwave drives applied to the individual qubits without flux tuning the qubit or coupler frequencies. In a shared-coupler architecture, however, the target-pair interaction must be selectively induced while suppressing unwanted effects on other qubits, and gates must be implemented for each direction.

A tunable coupler provides an additional control degree of freedom by allowing its frequency to be adjusted through an external magnetic flux.\cite{bialczak2011fast, yan2018tunable} The feasibility of controlled-phase gates has also been investigated in architectures where multiple qubits share a single tunable coupler.\cite{glaser2023controlled} Such architectures allow effective qubit interactions to be dynamically controlled through flux pulses applied to the coupler while retaining fixed-frequency data qubits. Selective gate implementation with a shared coupler therefore requires identifying pulse conditions that induce the desired interaction and quantitatively evaluating gate performance under those conditions.

In this work, we define an architecture in which three fixed-frequency transmon qubits are connected to one tunable coupler as a tunable tripod structure and investigate the feasibility of $\mathrm{CZ}$ gates driven by coupler flux modulation. Specifically, we aim to implement selective $\mathrm{CZ}$ gates for the three pairs $q_{0}\leftrightarrow q_{1}$, $q_{0}\leftrightarrow q_{2}$, and $q_{1}\leftrightarrow q_{2}$. We construct a system Hamiltonian that includes multiple energy levels of each component and calculate the time evolution of the full system under time-dependent flux pulses.

We apply virtual Z gates to correct single-qubit phases in the resulting operations and evaluate gate fidelity by comparison with an ideal $\mathrm{CZ}$ gate.\cite{marxer2026above} We further optimize the pulse parameters and analyze leakage. By optimizing the target-pair computational blocks formed from basis states in which both the coupler and spectator qubit are in their ground states, we obtain a maximum CZ gate fidelity of $99.43\%$. These results assess the feasibility of implementing CZ gates for each qubit pair through a shared coupler under the specified conditions.

\section{System Configuration}

The proposed architecture consists of three fixed-frequency qubits connected to a single tunable coupler. We refer to this architecture as a tunable tripod structure.

\subsection{Comparison with a Fixed-Frequency Coupler}
Previous studies investigated architectures with three fixed-frequency qubits connected to a single fixed-frequency coupler. Here, we replace the fixed-frequency coupler with a tunable coupler. In the absence of additional applied flux, this architecture can reproduce the fixed-frequency-coupler configuration studied previously. Applying additional flux also enables $\mathrm{iSWAP}$ and $\mathrm{CZ}$ gates, which are difficult to implement with a fixed-frequency coupler.

In systems composed of fixed-frequency qubits and fixed-frequency couplers, microwave drives applied to individual qubits serve as the primary control mechanism. These drives can implement $\mathrm{CX}$ gates through CR interactions. Because both the qubit and coupler frequencies are fixed, however, direct control over the effective qubit--qubit interaction and the associated resonance conditions is limited. As a result, implementing an $\mathrm{iSWAP}$ gate through the exchange interaction $|10\rangle \leftrightarrow |01\rangle$, or a $\mathrm{CZ}$ gate through a conditional interaction between $|11\rangle$ and a noncomputational state, is relatively difficult.

A tunable coupler, by contrast, provides an additional control degree of freedom because its frequency can be adjusted through an external magnetic flux. When the coupler is operated at a fixed flux-bias point, CR-based $\mathrm{CX}$ gates can be implemented using the conventional microwave-driven approach without directly tuning the qubit frequencies. Varying or modulating the flux applied to the coupler also dynamically controls its frequency and the effective interaction between a selected qubit pair. This allows resonance conditions to be selectively established for the desired pair, enabling $\mathrm{CZ}$ gates.

Fixed-frequency couplers avoid the additional flux-noise channel associated with flux tunability, at the cost of reduced interaction controllability. The proposed architecture introduces tunability only in the coupler while retaining fixed-frequency data qubits, with the aim of dynamically controlling different two-qubit interactions without directly flux tuning the qubits.

\subsection{Physical Model and System Design}
\begin{figure*}[t]
    \centering
    \includegraphics[width=0.8 \textwidth]{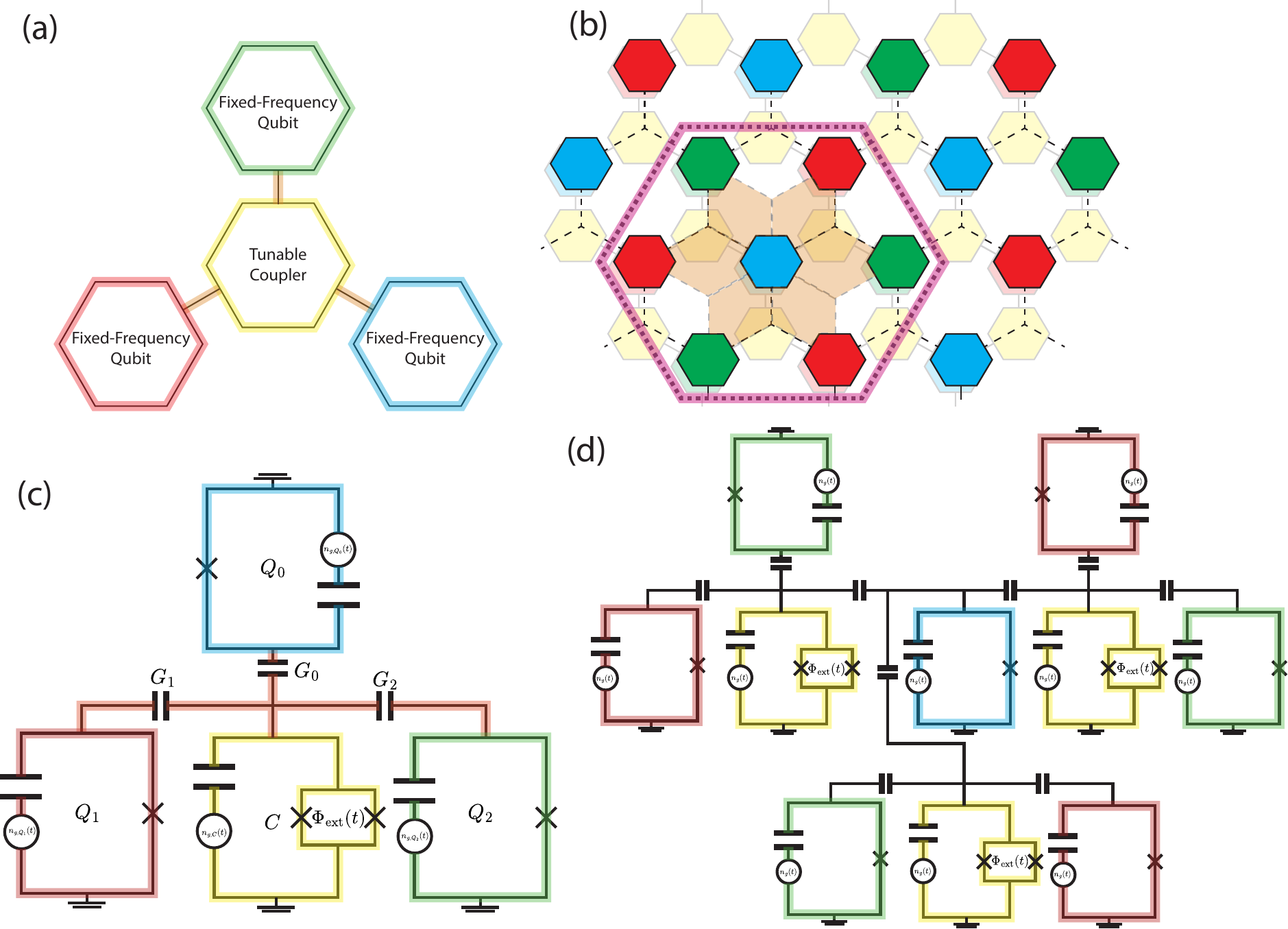}
    \caption{\label{fig:tripod_structure}
System architecture. (a) Proposed model with three fixed-frequency qubits connected to a single tunable coupler, shown as a yellow block. (b) An extended architecture based on the proposed structure. Red, blue, and green elements represent fixed-frequency qubits, and dashed lines indicate qubit connections mediated by the yellow couplers beneath them. Each qubit can connect to up to six other qubits. (c) Lumped-element circuit of the model in (a). Here, C denotes the coupler and $Q_{0},\;Q_{1},\;Q_{2}$ denote the fixed-frequency qubits. The tunable coupler $C$ includes a microwave pulse input and a DC-SQUID structure, which allows the relevant energy levels of the qubits to be brought closer together. Each qubit is capacitively connected to the coupler, with coupling strength $G_{i}$ for the $i$th qubit. (d) Lumped-element circuit corresponding to the pink hexagonal region in (b). The central blue fixed-frequency qubit is connected to three tunable couplers, providing connections to six qubits in total.
    }
\end{figure*}

The proposed system consists of three fixed-frequency qubits connected to one tunable coupler, as shown in Fig.~\ref{fig:tripod_structure}. Figure~\ref{fig:tripod_structure}(a) presents a schematic of the system: the red, green, and blue fixed-frequency qubits are directly connected to the yellow tunable coupler. An extension of this architecture is shown in Fig.~\ref{fig:tripod_structure}(b). The resulting honeycomb structure connects each qubit to up to three neighboring couplers. Because each coupler connects it to two additional qubits, a qubit at the center can have connectivity six. The lumped-element circuits corresponding to Figs.~\ref{fig:tripod_structure}(a) and \ref{fig:tripod_structure}(b) are shown in Figs.~\ref{fig:tripod_structure}(c) and \ref{fig:tripod_structure}(d), respectively. In Fig.~\ref{fig:tripod_structure}(c), $C$ denotes the tunable coupler and $Q_{0}$, $Q_{1}$, and $Q_{2}$ denote the qubits. Each qubit is capacitively connected to the coupler with coupling strength $G_{0}$, $G_{1}$, or $G_{2}$. Two-qubit gates are implemented through the external flux applied to coupler $C$, $\Phi_{\mathrm{ext}}(t)$, while individual qubit drives are represented by $n_{g}(t)$. Figure~\ref{fig:tripod_structure}(d) shows the lumped-element circuit for the pink hexagonal region in Fig.~\ref{fig:tripod_structure}(b). The central blue fixed-frequency qubit is connected to three tunable couplers, each linking it to two additional qubits, giving six qubit connections in total.

To describe the system dynamics, we introduce the total Hamiltonian. For computational convenience, we use natural units with $\hbar=1$. The total Hamiltonian is given by Eq.~\eqref{eq:total_hamiltonian}.
\begin{equation}\label{eq:total_hamiltonian}
    \hat{\mathcal{H}}=\hat{H}_{C}(t)+\sum_{i=0}^{2}\hat{H}_{Q,i}(t)+\hat{H}_{I}
\end{equation}
Here, $\hat{H}_{C}(t)$ is the coupler Hamiltonian, $\hat{H}_{Q,i}(t)$ is the Hamiltonian of the $i$th qubit, and $\hat{H}_{I}$ describes the interactions between the qubits and the coupler.

The tunable coupler consists of two Josephson junctions (JJs) connected in parallel to form a loop, rather than a single JJ. Its Hamiltonian $\hat{H}_{C}(t)$ is given by Eq.~\eqref{eq:tunable_coupler}.
\begin{equation}\label{eq:tunable_coupler}
    \hat{H}_{C}(t)=4E_{C,c}\hat{n}_{c}^{2}-E_{J,c}^{\mathrm{eff}}(\Phi_{\mathrm{ext}})\cos\hat{\phi}_{c}
\end{equation}
Here, $\hat{n}_{c}$ and $\hat{\phi}_{c}$ are the coupler number and phase operators, respectively, and satisfy the canonical commutation relation
\begin{equation*}
    [\hat{\phi}_{c},\hat{n}_{c}]=i
\end{equation*}

The parameter $E_{C,c}$ is the charging energy of the tunable coupler, and $E_{J,c}^{\mathrm{eff}}$ is the effective Josephson energy controlled by the external flux. For an asymmetric SQUID, this effective energy is given by Eq.~\eqref{eq:effective_josephson_energy}.
\begin{equation}\label{eq:effective_josephson_energy}
    E_{J,c}^{\mathrm{eff}}=E_{J,c}^{\Sigma}\left[ \cos^{2}\left(\frac{\varphi_{\mathrm{ext}}}{2}\right)+d^{2}_{c} \sin^{2}\left(\frac{\varphi_{\mathrm{ext}}}{2}\right) \right]^{1/2}
\end{equation}
Here, $\varphi_{\mathrm{ext}}$ is the reduced external flux, defined as
\begin{equation*}
    \varphi_{\mathrm{ext}}=\frac{2\pi \Phi_{\mathrm{ext}}}{\Phi_{0}},\quad \Phi_{0}=\frac{h}{2e}
\end{equation*}
The Josephson energies of the two junctions are denoted by $E_{J,c1}$ and $E_{J,c2}$. Their sum $E_{J,c}^{\Sigma}$ and the asymmetry parameter $d_{c}$ are defined as

\begin{equation*}
    E_{J,c}^{\Sigma}=E_{J,c1}+E_{J,c2},\quad d_{c}=\frac{E_{J,c1}-E_{J,c2}}{E_{J,c1}+E_{J,c2}}
\end{equation*}

The Hamiltonian $\hat{H}_{Q,i}(t)$ of each fixed-frequency qubit is given by Eq.~\eqref{eq:qubit_hamiltonian}.

\begin{equation}
    \label{eq:qubit_hamiltonian}
    \hat{H}_{Q,i}(t)=4E_{Q,C}(\hat{n}_{q,i}-n_{g}(t))^{2}-E_{Q,J}\cos\hat{\phi}_{q,i}
\end{equation}

The number operator $\hat{n}_{q,i}$ and phase operator $\hat{\phi}_{q,i}$ satisfy the same canonical commutation relation as those of the tunable coupler. The parameters $E_{Q,C}$ and $E_{Q,J}$ are the charging and Josephson energies, respectively. Each qubit has a microwave drive that can be used for single-qubit operations. In this work, however, gates are implemented by controlling the external flux of the tunable coupler. We therefore set $n_{g}(t)=0$, giving the qubit Hamiltonian used in the simulation:

\begin{equation}
    \label{eq:qubit_hamiltonian_real}
    \hat{H}_{Q,i}(t)=4E_{Q,C}\hat{n}_{q,i}^{2}-E_{Q,J}\cos\hat{\phi}_{q,i}
\end{equation}

Both the tunable coupler and fixed-frequency qubits operate in the transmon regime, $E_{C}\ll E_{J}$. This regime is less sensitive to noise than the charge-qubit regime, $E_{C}\gg E_{J}$, and is widely used in superconducting-qubit platforms.

Finally, we consider the interaction between the coupler and each qubit. The elements are connected capacitively, and their interaction is described, starting from the Jaynes--Cummings model, by a capacitive coupling between their charge operators, as given in Eq.~\eqref{eq:interaction_hamiltonian}.

\begin{equation}\label{eq:interaction_hamiltonian}
    \hat{H}_{I}=\sum_{i=0}^{2}G_{i}\hat{n}_{C}\hat{n}_{Q,i}
\end{equation}
This is a capacitive charge--charge interaction, where $G_{i}$ denotes the coupling strength between the coupler and the $i$th qubit.

\subsection{Multi-Level Numerical Simulation}
Previous studies analyzed leakage as the number of energy levels included in the simulation was increased. They found that leakage tends to converge once a sufficient number of levels is included, allowing quantitative evaluation without extending the Hilbert space to extremely high dimensions. The $\mathrm{CZ}$ gates considered here use interactions between the target-pair state $\vert 11\rangle$ and $\vert 20\rangle$ or $\vert 02\rangle$. To include transitions to these noncomputational levels, we truncate each element to the three-level basis $\{ \vert 0\rangle,\;\vert 1\rangle,\;\vert 2\rangle\}$. The basis of the full system is written as
\begin{equation}\label{eq:system_basis}
    \mathcal{S}=\vert k\rangle_{C}\otimes \bigotimes_{i=0}^{2}\vert m_{i}\rangle,\quad (k,m_{i}\in \{0,1,2\})
\end{equation}

\begin{figure*}[t]
    \centering
    \includegraphics[width=1 \textwidth]{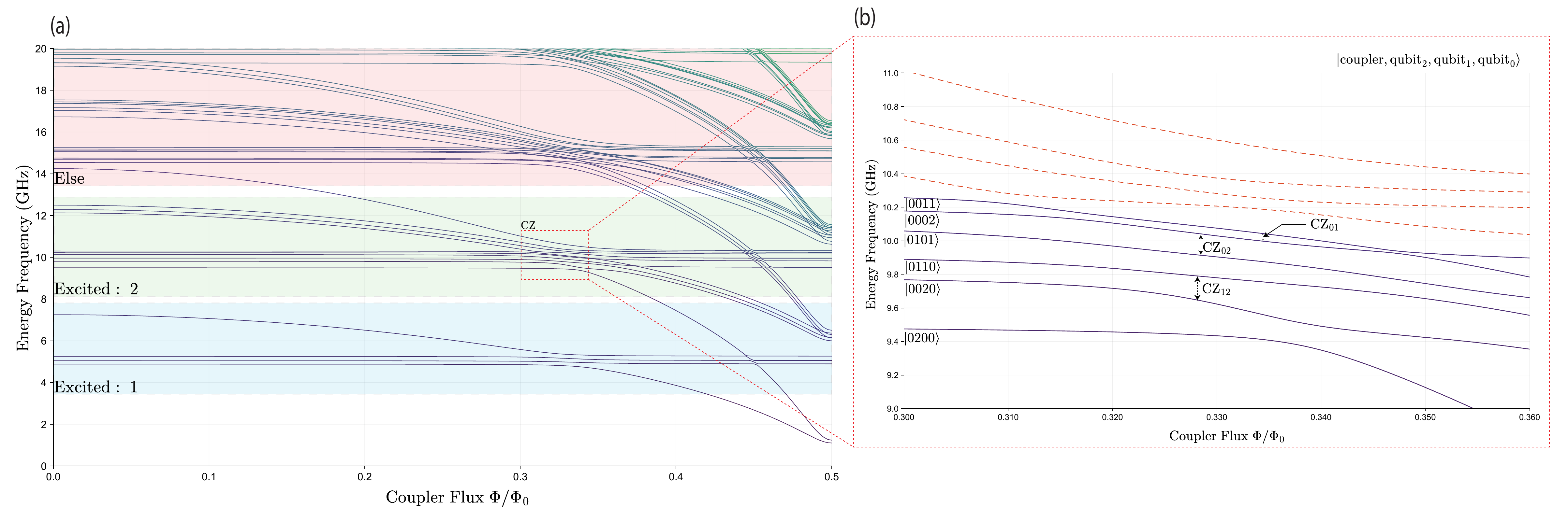}
    \caption{
    \label{fig:energy_spectrum}
Energy spectrum and candidate CZ operating points of the tunable tripod structure. (a) Variation of the energy levels as the flux applied to the coupler is swept over $[0,\pi]$. The spectrum provides information on the single- and two-excitation levels and identifies candidate CZ operating points. (b) Energy spectrum near the candidate CZ operating points. Solid lines indicate states with the coupler in its ground state, and red dashed lines indicate states with an excited coupler. For the CZ gate, the target-pair state $\vert ct\rangle$ is initialized in $\vert 11\rangle$, and the figure shows the interaction regions involving $\vert 20\rangle$ and $\vert 02\rangle$.
    }
\end{figure*}

The corresponding energy spectrum is shown in Fig.~\ref{fig:energy_spectrum}(a). Although up to 81 levels can be plotted, we focus on the single- and two-excitation sectors relevant to the present analysis.

\section{Pulse Protocol and Gate Parameter Optimization}

This section describes how pulse parameters are identified and optimized to implement $\mathrm{CZ}$ gates on the proposed hardware. In previous multi-qubit architectures with a single fixed-frequency coupler, pulse searches and optimization were performed using CR-based pulses and expectation values of computational-basis states. The present system instead uses a tunable coupler to implement a gate by adding phases to computational-basis states. Because its operating mechanism differs from that of fixed-frequency-coupler schemes, a different procedure is required to search for and optimize the pulse parameters.

\subsection{Pulse Protocol}
Gates on the proposed hardware are implemented by applying flux pulses to the coupler to adjust the system energy levels, rather than by applying microwave drive pulses to individual qubits. We use a parametric-resonance gate scheme, in which the relevant energy levels are brought closer together through the coupler, rather than the adiabatic pulse approach used in some CZ implementations.\cite{xu2020high, wang2019experimental, sete2024error} The flux applied to the coupler modifies the energy levels associated with the target pair, with the aim of establishing resonance between the two qubits.

The shape of the applied flux pulse plays an important role in gate implementation. The pulse protocol used in the simulation is given by Eq.~\eqref{eq:pulse_protocol}.

\begin{equation}\label{eq:pulse_protocol}
    \Phi_{\mathrm{ext}}(t) = \Phi_{\mathrm{off}} + \Phi_{\mathrm{ac}}\varepsilon(t)\cos(2\pi f_{d}t+\phi_{\mathrm{init}}).
\end{equation}
Here, $\Phi_{\mathrm{ext}}(t)$ is the external flux applied to the coupler, $\Phi_{\mathrm{off}}$ is a time-independent flux offset, and $\Phi_{\mathrm{ac}}$ is the flux modulation amplitude relative to that offset. The dimensionless pulse envelope is denoted by $\varepsilon(t)$, the modulation frequency by $f_d$, and the initial phase by $\phi_{\mathrm{init}}$. The angular frequency is defined as $\omega_d=2\pi f_d$. We distinguish the time variable $t$ from the total pulse duration $T$; $\Phi_{\mathrm{off}}$, $\Phi_{\mathrm{ac}}$, $f_d$, and $\phi_{\mathrm{init}}$ remain constant within each pulse.

Taking the pulse start time to be $t=0$, we define the envelope as

\begin{equation}
    \varepsilon(t) =
    \begin{cases}
        \sin^{2}\left(\dfrac{\pi t}{2t_{r}}\right), & 0\le t<t_{r}, \\
        1, & t_{r}\le t<T-t_{r}, \\
        \sin^{2}\left(\dfrac{\pi(T-t)}{2t_{r}}\right),
        & T-t_{r}\le t\le T, \\
        0, & \mathrm{otherwise}.
    \end{cases}
\end{equation}

Here, $T$ is the total pulse duration, and $t_r=t_{\mathrm{rise}}=t_{\mathrm{fall}}$ denotes the equal rise and fall times. Using the dimensionless ramp ratio $\rho$, we define

\begin{equation}
    t_{r} = \rho T, \qquad 0<\rho<0.5
\end{equation}
as the rise and fall time.

This sinusoidal flat-top flux pulse modulates the external flux applied to the coupler, thereby changing the coupler frequency and the effective qubit interactions mediated by it. We define a bare state as a tensor product of the eigenstates of the individual elements with interelement coupling neglected. A dressed state is an eigenstate of the full, coupled Hamiltonian at a fixed flux. Omitting the coupler and spectator labels, the target-pair states $|10\rangle$, $|01\rangle$, and $|11\rangle$ are computational bare states, whereas $|20\rangle$ and $|02\rangle$ are noncomputational bare states. A dressed state is generally a superposition of multiple bare-state components. In the simulation, flux modulation establishes a resonance condition for the target transition. In bare-state notation, the $\mathrm{CZ}$ gate uses an interaction between states corresponding to $|11\rangle$ and either $|20\rangle$ or $|02\rangle$. The $\mathrm{CZ}$ pulse is designed to return the population to the computational subspace at the end of the gate while producing a conditional phase of $\pi$.

\subsection{Population Dynamics and Gate Fidelity}
In this work, the $\mathrm{CZ}$ gate is implemented by controlling the flux of the tunable coupler. Its ideal matrix is
\begin{equation}\label{eq:opt_operator}
    \mathrm{CZ}=\begin{bmatrix}
        1 & 0 & 0 & 0 \\
        0 & 1 & 0 & 0 \\
        0 & 0 & 1 & 0 \\
        0 & 0 & 0 & -1
    \end{bmatrix}
\end{equation}

In the proposed system, each qubit is connected to the coupler, which protects the qubit frequencies. To induce an interaction between two selected qubits, flux is applied to the coupler to adjust the energy levels of the corresponding two-qubit system. The energy diagram is shown in Fig.~\ref{fig:energy_spectrum}(b).

Figure~\ref{fig:energy_spectrum}(b) shows the region of the two-excitation spectrum in Fig.~\ref{fig:energy_spectrum}(a) identified as a candidate for the $\mathrm{CZ}$ gate. Here, the total number of excitations across the coupler and qubits is two. The $\mathrm{CZ}$ gate involves interactions among $\vert 11\rangle \leftrightarrow \vert 20\rangle \leftrightarrow \vert 02\rangle$. Compared with the single-excitation sector, the increased number of excitations produces a more complex spectrum, making the positions of the relevant energy levels an important consideration for $\mathrm{CZ}$ implementation.

\begin{figure*}[t]
    \centering
    \includegraphics[width=0.9 \textwidth]{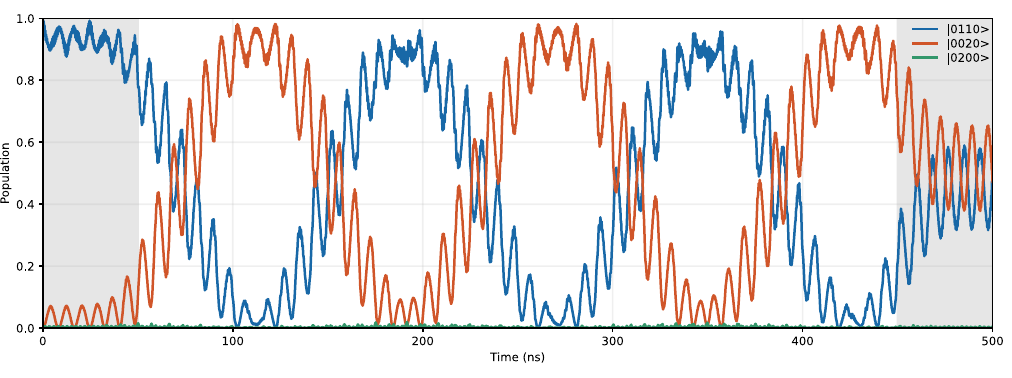}
    \caption{
    \label{fig:gate_population}
State populations under the flux applied to the coupler. The populations of $\vert 0110\rangle$, $\vert 0020\rangle$, and $\vert 0200\rangle$ are shown over $500\;ns$. The rise and fall times are both 50 $ns$, as indicated by the gray regions. An interaction is observed with $\vert 0020\rangle$, whereas no appreciable interaction occurs with $\vert 0200\rangle$. This illustrates the importance of selecting the bare state and the dressed state that mediates its interaction when implementing a $\mathrm{CZ}$ gate on $\vert 11\rangle$.
    }
\end{figure*}

To examine this energy exchange, we applied flux modulation to the coupler and monitored the state populations. Because the $\mathrm{CZ}$ gate uses interactions between computational and noncomputational states, Fig.~\ref{fig:gate_population} shows the populations of the states corresponding to $\vert 11\rangle$, $\vert 20\rangle$, and $\vert 02\rangle$ for the $q_0\leftrightarrow q_1$ pair.

Figure~\ref{fig:gate_population} shows population exchange between $\vert 0110\rangle$ and $\vert 0020\rangle$ under coupler flux modulation, while the population of $\vert 0200\rangle$ remains small.

Observing population exchange alone does not guarantee high $\mathrm{CZ}$ gate fidelity. The corresponding pulse conditions were therefore used as starting points for identifying initial parameters that induce the target transition.

In the simulation, we performed a parameter sweep over $(\Phi_{\mathrm{ac}},f_{d},T)$ with no flux offset ($\Phi_{\mathrm{off}}=0$) and initial phase $\phi_{\mathrm{init}}=0$. We denote the time-evolution operator of the full Hamiltonian for each pulse condition by $\hat{U}_{\mathrm{full}}(T)$. From this operator, we select the four computational-basis states corresponding to the target pair of the $\mathrm{CZ}_{ij}$ gate as the input and output basis to construct a $4\times 4$ computational block $\hat{U}_{\mathrm{comp}}$. Denoting the selected full-system basis states by $\vert\psi_{m}\rangle$, we write

\begin{equation}\label{eq:computational_block}
    (U_{\mathrm{comp}})_{nm}
    =\langle \psi_{n}\vert \hat{U}_{\mathrm{full}}(T)\vert \psi_{m}\rangle ,
    \quad m,n=0,\ldots,3.
\end{equation}
This block is not renormalized. Because population can leave the selected subspace, it is generally nonunitary.

Single-qubit phases of the target pair are corrected using virtual Z gates.\cite{mckay2017efficient} Denoting the target qubits by $i,j$, the correction operator is written as
\begin{equation}\label{eq:virtual_z}
    R_{Z}^{(\mathbf{v})}(\mathbf{\theta})=\bigotimes_{k=0}^{2}R_{z}^{(k)}(\theta_{k})
\end{equation}
Here, $R_{Z}^{(k)}(\theta)$ represents an additional virtual Z gate with phase angle $\theta$ on the $k$th qubit. With this correction, target-pair gate fidelity is evaluated and optimized within the computational block corresponding to the pair on which the gate acts. The corrected computational block is defined as
\begin{equation}\label{eq:total_unitary}
    \hat{U}_{\mathrm{final}}
    =\hat{R}_{Z}^{(ij)}(\theta_i,\theta_j)\hat{U}_{\mathrm{comp}}
\end{equation}
We denote the ideal target-pair $\mathrm{CZ}$ gate by $\hat{U}_{\mathrm{ideal}}$ and calculate the average gate fidelity by comparison with the corrected block:
\begin{equation}\label{eq:fidelity}
    \mathcal{F}_{\mathrm{avg}}
    =\frac{\mathrm{Tr}(\mathcal{M}^{\dagger}\mathcal{M})+|\mathrm{Tr}(\mathcal{M})|^{2}}{d(d+1)}.
\end{equation}
where
\begin{equation}\label{eq:operator_difference}
    \mathcal{M}=\hat{U}_{\mathrm{ideal}}^{\dagger}\hat{U}_{\mathrm{final}}
\end{equation}
is the overlap matrix comparing the implemented operation with the ideal gate, and $d=4$ is the dimension of the target-pair computational subspace.

\subsection{Gate Parameter Optimization}

Using the conditions obtained from the parameter sweep as initial values, we optimize the parameters with the Nelder--Mead algorithm. The optimization objective is based on the average gate fidelity of the selected target-pair computational block after virtual Z correction. The parameter vector specifying the pulse conditions and virtual Z correction angles is

\begin{equation}\label{eq:optimizer_parameters}
    \boldsymbol{\lambda}
    =\left(\Phi_{\mathrm{ac}},f_d,T,\rho,
    \phi_{\mathrm{init}},\theta_i,\theta_j\right)
\end{equation}
Here, $\rho$ is the envelope ramp ratio, while $\theta_{i}$ and $\theta_{j}$ are the rotation angles, rather than the rotation operators, of the target-pair virtual Z gates. The final parameters for each gate, including the modulation frequency $f_d$, are presented in the next section.

The optimization loss function is defined as
\begin{equation}\label{eq:opt_loss_function}
    J(\boldsymbol{\lambda})
    =1-\mathcal{F}_{\mathrm{avg}}(\boldsymbol{\lambda})
\end{equation}
This objective incorporates population loss from the selected computational subspace as well as unwanted state transitions and phase errors within the subspace.

\section{Results}
To assess the feasibility of the proposed tunable tripod structure, we specify the hardware parameters, present the flux parameters used to implement the $\mathrm{CZ}$ gates, and report the fidelities of $\mathrm{CZ}_{01}$, $\mathrm{CZ}_{02}$, and $\mathrm{CZ}_{12}$.

\subsection{Hardware Specification}
The tunable tripod structure comprises three fixed-frequency transmon qubits connected to a single tunable coupler. The coupler and qubit parameters used in the simulation are summarized in Table~\ref{tab:hardware_specification}.

\begin{table*}[t]
\caption{
\label{tab:hardware_specification}
Hardware parameters used in the numerical simulation of the tunable tripod structure. Hamiltonian parameters are expressed in frequency units obtained by dividing angular frequencies by $2\pi$. Here, $E_C$ and $E_J$ are the charging and Josephson energies, respectively, and $G_{C,i}$ is the capacitive coupling strength between the coupler and the $i$th qubit.
}
\begin{ruledtabular}
\begin{tabular}{ccccc}
\textrm{Parameter} & \textrm{Coupler} & \textrm{Qubit$_0$} & \textrm{Qubit$_1$} & \textrm{Qubit$_2$} \\
\colrule

$E_C/2\pi$ (GHz) & 0.229 & 0.297 & 0.3204 & 0.3010 \\
$E_J/2\pi$ (GHz) & 29.1 & 11.1837 & 12.114 & 11.412 \\
$f_{01}$ (GHz) & 7.0684 & 4.8356 & 5.2280 & 4.9188 \\
$G_{C,i}/2\pi$ (GHz) & - & 0.04 & 0.04 & 0.04 \\

\end{tabular}
\end{ruledtabular}
\end{table*}

The coupler is an asymmetric DC-SQUID with two Josephson junctions connected in parallel. Its charging energy is $E_{C,c}/2\pi=0.229\;\mathrm{GHz}$, and the sum of the two Josephson energies is $E_{J,c}^{\Sigma}/2\pi=29.1\;\mathrm{GHz}$, as listed in Table~\ref{tab:hardware_specification}. The junction Josephson-energy ratio is set to $E_{J,c1}/E_{J,c2}=1.04$, corresponding to an asymmetry parameter $d_c=(E_{J,c1}-E_{J,c2})/(E_{J,c1}+E_{J,c2}) \simeq 0.0196$.

Each fixed-frequency transmon qubit consists of a capacitor and a single Josephson junction. To obtain distinct transition frequencies near $5\;\mathrm{GHz}$, the charging energies $E_{C,Q_i}/2\pi$ of $Q_0$, $Q_1$, and $Q_2$ are set to 0.297, 0.3204, and $0.3010\;\mathrm{GHz}$, respectively. The corresponding Josephson energies $E_{J,Q_i}/2\pi$ are 11.1837, 12.114, and $11.412\;\mathrm{GHz}$. All capacitive couplings between the coupler and qubits are set to $G_{C,i}/2\pi=0.04\;\mathrm{GHz}$.

After setting the charging and Josephson energies, we numerically diagonalize the Hamiltonian of each element to obtain its eigenenergies. The reference coupler frequency is calculated at $\Phi_{\mathrm{ext}}=0$, and the transition frequency $f_{01}$ of each element is obtained from the energy difference between the ground and first excited states. The reference coupler frequency in Table~\ref{tab:hardware_specification} is $7.0684\;\mathrm{GHz}$, and the frequencies of $Q_0$, $Q_1$, and $Q_2$ are 4.8356, 5.2280, and $4.9188\;\mathrm{GHz}$, respectively. The pairwise frequency differences $|f_{01}^{(i)}-f_{01}^{(j)}|$ for $q_0\leftrightarrow q_1$, $q_0\leftrightarrow q_2$, and $q_1\leftrightarrow q_2$ are consequently 392.4, 83.2, and $309.2\;\mathrm{MHz}$, respectively.

\subsection{Pulse Protocol}

Table~\ref{tab:pulse_parameters} lists the final parameters of the flux pulse in Eq.~\eqref{eq:pulse_protocol} and the target-pair virtual Z corrections. All gates use $\Phi_{\mathrm{off}}=0$, and the notation follows Section III. The flux amplitude is normalized by the flux quantum $\Phi_{0}$ and reported as $\Phi_{\mathrm{ac}}/\Phi_{0}$.

\begin{table*}[t]
\caption{\label{tab:pulse_parameters}
Final pulse parameters and virtual Z correction angles used to implement each $\mathrm{CZ}_{ij}$ gate in the tunable tripod structure. Here, $T$ is the total pulse duration, and the flux modulation amplitude is normalized as $\Phi_{\mathrm{ac}}/\Phi_0$. The dimensionless ramp ratio is $\rho$, and the modulation frequency is $f_d=\omega_d/(2\pi)$. Equal rise and fall times are used, with $t_r=t_{\mathrm{rise}}=t_{\mathrm{fall}}=\rho T$. The initial phase is $\phi_{\mathrm{init}}$, while $\theta_{i}$ and $\theta_{j}$ are the virtual Z rotation angles applied to target qubits $i,j$.
}
\begin{ruledtabular}
\begin{tabular}{cccccccc}
Gate & $T$ (ns) & $\Phi_{\mathrm{ac}}/\Phi_{0}$ & $\rho$ & $f_{d}$ (GHz) & $\phi_{\mathrm{init}}$ (rad) & $\theta_{i}$ (rad) & $\theta_{j}$ (rad) \\
\colrule
$\mathrm{CZ}_{01}$ & 123.53 & 0.3149 & 0.1088 & 0.3152 & 0.1865 & 2.2899 & 2.2899\\
$\mathrm{CZ}_{02}$ & 147.57 & 0.3489 & 0.099 & 0.3030 & 0.1048 & -0.6983 & 0.1674\\
$\mathrm{CZ}_{12}$ & 117.00 & 0.3419 & 0.1027 & 0.3306 & -0.5943 & -1.7768 & 0.5428\\
\end{tabular}
\end{ruledtabular}
\end{table*}

The pulse duration $T$ and ramp ratio $\rho$ determine the temporal structure of the envelope, with rise and fall times of $\rho T$. The flux modulation amplitude $\Phi_{\mathrm{ac}}$ and modulation frequency $f_d$ control the interaction associated with the target transition. The reported $f_d$ is the final modulation frequency obtained by searching around the resonance condition of the target transition between the coupled dressed-state levels defined in Section III and refining it based on gate fidelity. The parameter $\phi_{\mathrm{init}}$ is the initial flux-pulse phase, and $\theta_i$ and $\theta_j$ are the virtual Z correction angles applied to the target pair after the pulse.

For each pulse, we calculate the full-Hamiltonian time evolution and extract the $4\times 4$ computational block defined in Eq.~\eqref{eq:computational_block}. We compare the target-pair virtual-Z-corrected operator $\hat{U}_{\mathrm{final}}$ with the ideal $\mathrm{CZ}$ gate $\hat{U}_{\mathrm{ideal}}$ and evaluate $\mathcal{F}_{\mathrm{avg}}$ using Eq.~\eqref{eq:fidelity} with $d=4$. The block $U_{\mathrm{comp}}$ used in the optimization is constructed with the coupler and spectator qubit in their ground states. Because the computational block is not separately renormalized, $\hat{U}_{\mathrm{final}}$ is generally nonunitary. The average gate fidelities are reported in Table~\ref{tab:fidelity}.

\begin{table}[t]
\caption{
Average CZ gate fidelity $\mathcal{F}_{\mathrm{avg}}$ evaluated for the selected target-pair computational block. The fidelity is calculated using Eq.~\eqref{eq:fidelity} after target-pair virtual Z correction. \label{tab:fidelity}
}
\begin{ruledtabular}
\begin{tabular}{cccc}
&$\mathrm{CZ}_{01}$ & $\mathrm{CZ}_{02}$ & $\mathrm{CZ}_{12}$ \\
\colrule

Fidelity ($\mathcal{F}_{\mathrm{avg}}$) & 0.9943 & 0.9834 & 0.9907 

\end{tabular}
\end{ruledtabular}
\end{table}

\subsection{Leakage Error and Error Budget}
The $\mathrm{CZ}$ gates in the tunable-coupler architecture are implemented by inducing interactions between the target qubits through the flux applied to the coupler. An ideal $\mathrm{CZ}$ gate multiplies the target-pair state $\vert 11\rangle$ by a phase factor of $-1$, with population returning to the computational subspace at the end of the gate.

We define leakage $\mathcal{L}$ as the average population remaining outside the selected computational subspace at the end of the gate. Using the same four-dimensional computational subspace as in the fidelity evaluation, we calculate it as
\begin{equation}\label{eq:leakage}
    \begin{aligned}
    \mathcal{L}&=1-\frac{1}{d}\sum_{m,n=0}^{d-1}\left|\langle \psi_{n}\vert \hat{U}_{\mathrm{full}}(T)\vert \psi_{m}\rangle \right|^2\\
    &=1-\frac{1}{d}\mathrm{Tr}\left(\hat{U}_{\mathrm{comp}}^{\dagger}\hat{U}_{\mathrm{comp}}\right),\qquad d=4.
    \end{aligned}
\end{equation}
Here, $\vert \psi_{m}\rangle$ denotes one of the full-system basis states selected in Eq.~\eqref{eq:computational_block}. This definition includes not only coupler excitation and occupation of noncomputational qubit levels, but also departures of the spectator from $\vert 0\rangle$ that take the system outside the selected four-dimensional subspace. It must therefore be distinguished from leakage defined with respect to the full three-qubit computational space. Virtual Z correction does not change the block population and hence does not affect $\mathcal{L}$.

The resulting values of $\mathcal{L}$ for each gate are listed in Table~\ref{tab:leakage}.

\begin{table}[t]
\caption{
\label{tab:leakage} Target-pair subspace leakage $\mathcal{L}$ of the implemented $\mathrm{CZ}$ gates. Using the four-dimensional computational subspace in which both the coupler and spectator qubit are in their ground states, we evaluate the average population remaining outside this subspace at the end of the gate using Eq.~\eqref{eq:leakage}. Values are reported as probabilities rather than percentages.
}

\begin{ruledtabular}
\begin{tabular}{cccc}
 & $\mathrm{CZ}_{01}$ & $\mathrm{CZ}_{02}$ & $\mathrm{CZ}_{12}$ \\
\colrule
Leakage ($\mathcal{L}$) & 0.0047 & 0.0139 & 0.0086 \\
\end{tabular}
\end{ruledtabular}
\end{table}

Tables~\ref{tab:fidelity} and \ref{tab:leakage} report $\mathcal{F}_{\mathrm{avg}}$ and $\mathcal{L}$ evaluated for the same target-pair computational subspace. For $\mathrm{CZ}_{01}$, for example, the selected basis states are $\vert0000\rangle$, $\vert 0010\rangle$, $\vert 0100\rangle$, and $\vert 0110\rangle$. Average fidelity measures agreement with the ideal gate, whereas leakage measures the extent to which population returns to the selected subspace. Because the objective $J=1-\mathcal{F}_{\mathrm{avg}}$ includes leakage as well as state-transition and phase errors within the subspace, $J$ and $\mathcal{L}$ should not be interpreted as the same error measure.

\section{Discussion} 

We numerically implemented $\mathrm{CZ}$ gates to assess the feasibility of an architecture comprising three fixed-frequency qubits connected to a single tunable coupler. Selecting the relevant basis states from the full unitary obtained through full-Hamiltonian simulation and flux-pulse optimization is important for implementing the target two-qubit gate.

Because the proposed system uses fixed-frequency qubits, direct frequency adjustment to bring qubits into or out of interaction is limited. Introducing tunable qubits could provide additional frequency control and may improve gate fidelity. It is also important to examine the qubit outside the target pair when gate performance is evaluated using only selected basis states. We refer to this nonparticipating qubit as the spectator qubit. Its final-state probabilities after the implemented gates are shown in Table~\ref{tab:cz_spectator_qubit}.

\begin{table*}[t]
\caption{
\label{tab:cz_spectator_qubit} Spectator-qubit state probabilities for the $\mathrm{CZ}$ gates. The table shows the spectator computational-state probabilities after applying each $\mathrm{CZ}$ gate to the eight computational-basis input states.
}
\begin{ruledtabular}
\begin{tabular}{c|cccccc}
     &\multicolumn{2}{c}{$\mathrm{CZ}_{01}$} & \multicolumn{2}{c}{$\mathrm{CZ}_{02}$} & \multicolumn{2}{c}{$\mathrm{CZ}_{12}$} \\
    &$q_{2}(P_{0})$ &$q_{2}(P_{1})$ &$q_{1}(P_{0})$ &$q_{1}(P_{1})$&$q_{0}(P_{0})$ &$q_{0}(P_{1})$\\

\colrule
$\vert 000\rangle$ & $\mathbf{0.999993}$ & 0.000007 & $\mathbf{0.999990}$ & 0.000009 & $\mathbf{0.999995}$ & 0.000005 \\
$\vert 001\rangle$ & 0.000317 & 0.999667 & $\mathbf{0.997740}$ & 0.002257 & $\mathbf{0.998856}$ & 0.001143 \\
$\vert 010\rangle$ & $\mathbf{0.999943}$ & 0.000055 & 0.025341 & 0.974612 & $\mathbf{0.998731}$ & 0.001265 \\
$\vert 011\rangle$ & 0.135440 & 0.861415 & 0.122280 & 0.877353 & $\mathbf{0.996651}$ & 0.003346 \\
$\vert 100\rangle$ & $\mathbf{0.999734}$ & 0.000266 & $\mathbf{0.998807}$ & 0.001182 & 0.017588 & 0.982387 \\
$\vert 101\rangle$ & 0.131360 & 0.868479 & $\mathbf{0.995457}$ & 0.004463 & 0.051167 & 0.948637 \\
$\vert 110\rangle$ & $\mathbf{0.999215}$ & 0.000775 & 0.046612 & 0.876339 & 0.125374 & 0.868322 \\
$\vert 111\rangle$ & 0.235878 & 0.740892 & 0.150032 & 0.820973 & 0.384297 & 0.609035 \\
\end{tabular}
\end{ruledtabular}
\end{table*}

For the implemented $\mathrm{CZ}$ gates, a spectator initially in $\vert 0\rangle$ remains in its ground state with a probability of approximately 0.99, whereas a spectator initially in $\vert 1\rangle$ can remain in $\vert 1\rangle$, transfer population to $\vert 0\rangle$ or leave the computational basis. This behavior appears as leakage associated with the $\mathrm{CZ}$ gate implementation mechanism. For the target $\mathrm{CZ}01$ gate, the relevant prepared state is $\vert 0\rangle_{c}\times\vert 110\rangle_{q}$. When the spectator is in state 1, the qubit states $\vert 0\rangle_{c}\times\vert 101\rangle_{q}$ or $\vert 0\rangle_{c}\times\vert 011\rangle_{q}$ can instead be prepared. These states can interact directly with $\vert 0\rangle_{c}\times\vert 200\rangle_{q}$ and $\vert 0\rangle_{c}\times\vert 020\rangle_{q}$, which are the dressed states associated with $\vert 0\rangle_{c}\times\vert 110\rangle_{q}$. The spectator state is therefore an important consideration when implementing $\mathrm{CZ}$ gates through a single shared coupler.

The observed changes in spectator population show that flux modulation of the shared coupler can affect states outside the target pair. If these additional interactions can be selectively controlled, they may also be exploited to implement multi-qubit gates. In particular, in the model with three levels per qubit, the following states can participate in interactions when the coupler is in its ground state and the three qubits have a total of three excitations:
\begin{equation*}
    \mathcal{S}_{3}=
    \left\{
    \vert111\rangle,\,
    \vert102\rangle,\,
    \vert120\rangle,\,
    \vert201\rangle,\,
    \vert021\rangle,\,
    \vert012\rangle,\,
    \vert210\rangle
    \right\}.
\end{equation*}
The actual transitions and coupling strengths among these states depend on their energy differences and on the pulse conditions.

As the total excitation number increases, the level structure and possible transition paths become more complex. It is therefore important to selectively induce the target interaction through parametric resonance while suppressing unwanted transitions. A three-qubit gate must also return population to the computational subspace at the end of the gate and generate the intended relative phase. Future work should examine tunable qubits as a means of adjusting the level spacings and analyze how this additional control affects multi-qubit gate implementation and fidelity.

To explore an extension to multi-qubit gates, we performed a preliminary $\mathrm{CCZ}$ analysis using the parameter-sweep results originally obtained for $\mathrm{CZ}$ implementation. An ideal $\mathrm{CCZ}$ gate applies a phase factor of $-1$ only to $\vert 111\rangle$ in the three-qubit computational basis. Its ideal operator in the system computational basis is

\begin{equation}
    \label{eq:ccz_target}
    U_{\mathrm{CCZ}}=\mathrm{diag}(1,1,1,1,1,1,1,-1).
\end{equation}

We use the same fidelity evaluation as for the $\mathrm{CZ}$ gates, but with $d=8$ for the full three-qubit computational block and the ideal $\mathrm{CCZ}$ gate as the target. The highest corrected fidelity among the stored results is $\mathcal{F}_{\mathrm{avg}}^{\mathrm{CCZ}}=0.84149$. The corresponding pulse amplitude is $\Phi_{\mathrm{ac}}/\Phi_0=0.305$, the modulation frequency is $f_d=0.355\;\mathrm{GHz}$, the pulse duration is $T=145\;\mathrm{ns}$, and the ramp ratio is $\rho=0.1$. Both the flux offset and initial phase are zero. The virtual Z correction angles are $(\theta_0,\theta_1,\theta_2)=(-0.5059,\,1.3843,\,0.5494)\;\mathrm{rad}$. The leakage relative to the eight-dimensional subspace in which the coupler is in its ground state and all three qubits are in their computational basis is $L_{\mathrm{leak}}^{(8)}=0.1151$, or approximately $11.51\%$. Its reference subspace differs from that of $\mathcal{L}$ in Section IV, and transitions within the full three-qubit computational space do not contribute to $L_{\mathrm{leak}}^{(8)}$. These results are a preliminary analysis of existing parameter-sweep data, rather than the outcome of additional optimization targeting $\mathrm{CCZ}$.

These results indicate that extending shared-coupler flux modulation to multi-qubit gates requires controlling both the target three-qubit phase and the return of population to the computational space. The present results constitute a preliminary search using a restricted pulse form and a finite parameter grid; the phase correction likewise does not guarantee a global optimum. The highest fidelity found here should therefore not be interpreted as a fundamental performance limit of the architecture. Future work should directly optimize the pulse for $\mathrm{CCZ}$ fidelity and separately analyze unwanted two-qubit conditional phases, state transitions within the computational space, and leakage to noncomputational levels. Convergence checks with respect to the number of energy levels and the time-integration accuracy will also be needed to quantitatively assess the feasibility of extending the approach to multi-qubit gates.

\section{Conclusion}
We numerically investigated the feasibility of two-qubit gates in a tunable tripod structure consisting of three fixed-frequency transmon qubits connected to a single tunable coupler. Using a common flux-pulse shape applied to the coupler, we optimized its parameters for each target pair to construct $\mathrm{CZ}$ gates for all three qubit pairs. We calculated the gate dynamics using the full Hamiltonian and evaluated performance through gate fidelity and leakage.

The $\mathrm{CZ}$ gates evaluated within the selected computational blocks have fidelities ranging from 0.9834 to 0.9943 across the three pairs. These results show that pairwise $\mathrm{CZ}$ gates can be constructed through flux control of a shared coupler while retaining fixed-frequency qubits. The spectator-population analysis also shows that state preservation depends on the applied gate and initial computational-basis state. Evaluating a shared-coupler architecture therefore requires considering both target-pair gate fidelity and the dynamics of the full system, including the spectator.

This study examines both gate implementation and spectator effects in a three-qubit architecture sharing a single coupler. Future improvements may be achieved through pulse optimization that accounts for spectator effects and through adjustment of the hardware parameters. Analyses including relaxation, dephasing, and flux noise are also required to assess performance under experimental operating conditions. Building on these validations, future work should explore increased connectivity by connecting more qubits to a single coupler, as well as extensions to multi-qubit gates.

\section{Acknowledgement}
This work is supported by the Creation of the Quantum Information Science R$\&$D Ecosystem (Grant No. RS-2023-NR068116) through the National Research Foundation of Korea (NRF) funded by the Korean government (Ministry of Science and ICT).

%

\end{document}